\documentclass[twocolumn,showpacs,preprintnumbers,prb,fleqn,floatfix]{revtex4}

\usepackage{graphicx}
\usepackage{dcolumn}
\usepackage{bm}

\begin{document}
\title{{ Vortex-like excitations in a non-superconducting single-layer
compound Bi$_{2+x}$Sr$_{2-x}$CuO$_{6+\delta }$ single
crystal in high magnetic fields }}

\author{S. I. Vedeneev$^{1,2}$ and D. K. Maude$^{1}$}

\affiliation{ $^1$ Grenoble High Magnetic Field Laboratory, Centre
National de la Recherche Scientifique, B.P. 166, F-38042 Grenoble
Cedex 9, France \\
$^2$ P.N. Lebedev Physical Institute, Russian Academy of Sciences,
119991 Moscow, Russia}

\date{\today }

\begin{abstract}
The in-plane $\rho _{ab}(H,T)$ and the out-of-plane $\rho _c(H,T)$
magneto-transport in magnetic fields up to 28~T has been
investigated in high quality non-superconducting (down to 20 mK)
La-free Bi$_{2+x}$Sr$_{2-x}$CuO$_{6+\delta }$ single crystal. By
measuring the angular dependence of the in-plane and out-of-plane
magnetoresistivities at temperatures from 1 K down to $30$ mK, we
present evidence for the presence of vortex-like excitations in a
non-superconducting cuprate in the insulating state. Such
excitations have previously been observed by the detection of a
Nernst signal in superconducting cuprates at $T>T_c$ in magnetic
fields.
\end{abstract}

 \maketitle

\section{Introduction}

Originating from the pioneering experiments of Uemura \textit{et
al.},\cite{Uemura} there is now a general consensus concerning the
determining factor for the critical temperature ($T_c$) in high-
$T_c$ superconductors. In Ref.[\onlinecite{Uemura}] it was shown
that $T_c$ is proportional to the zero-temperature superconducting
carrier density for a wide range of underdoped materials. This
correlation is a consequence of the proximity to the Mott
transition.\cite{Orenstein} In conventional superconductors, the
destruction of superconductivity begins with the breakup of
electron pairs. However, in cuprates with increasing temperature,
thermal excitations will destroy the ability of the superconductor
to carry a supercurrent whereas the pairs can continue to
exist.\cite{Xu}

At present, there is growing evidence that the transition out of
the superconducting state is caused by the proliferation of
vortices, which destroy long-range phase coherence. The detection
of a large Nernst signal above $T_c$ has provided evidence for the
vortex scenario.\cite{Corson, Xu, Capan, Wang02, Wang03} Recently,
Sandu \textit{et al.} \cite{Sandu} have shown that the measured
in-plane angular dependence of the magnetoresistance on
Y$_{1-x}$Pr$_x$Ba$_2$Cu$_3$O$_{7-\delta}$ single crystals with 35
K $\le T_c \le$ 92 K is consistent with a flux-flow type
contribution. This is again an indication of the presence of
vortex-like excitations above $T_c$ in the pseudogap region.

However, in Ref. [\onlinecite{Tan}], it has been suggested that
the large Nernst signal observed above $T_c$ is due to
superconducting fluctuations in the normal state. Very recently,
Alexandrov and Zavaritsky \cite {Alexandrov} calculated the
expected Nernst signal in disordered conductors and showed that a
strong Nernst signal is unrelated to vortices or a superconducting
pair scenario. They found instead, that the Nernst signal could
arise from the interference of itinerant and localized-carrier
contributions to the thermomagnetic transport.

Therefore, whether these phenomena are a result of the presence of
vortices above the zero-field critical temperature remains an open
question. Our magnetoresistance investigation of a
non-superconducting single crystal at temperature down to 30~mK
should help to distinguish between these different points of view
because magnetoresistance is a potentially incisive probe of the
hole dynamics. In this paper we present, to our knowledge, the
first evidence for vortices in a non-superconducting cuprate from
angular magnetoresistance measurements on La-free, high quality
Bi$_{2+x}$Sr$_{2-x}$CuO$_{6+\delta }$ (Bi2201) single crystals.
Transport and magnetotransport in non-superconducting Bi2201
crystals have been investigated long ago (see e.g. Refs.
[\onlinecite{Martin, Fiory, Forro, Jing}]), however,
magnetoresistance and the angular dependence of the
magnetoresistance at mK temperatures was not studied.

In previous measurements \cite{Vedeneev04} we have investigated
the in-plane $\rho _{ab}(H,T)$ and the out-of-plane $\rho _c(H,T)$
magneto-transport in magnetic fields up to 28~T in a series of
superconducting Bi2201 single crystals over a wide doping range
and over a wide range of temperatures down to $40$ mK. The
$T_c$(midpoint) values of the crystals lay in the region
$2.3-9.6$~K. With decreasing carrier concentration per Cu atom
($p$), going from the overdoped ($p=0.2$) to the underdoped
($p=0.12$) regimes, a crossover from a metallic to an insulating
behavior of $\rho _{ab}(T)$ was observed in the low temperature
normal state, resulting from a disorder induced metal insulator
transition. Note that, throughout this paper, by insulating phase
we simply mean that the resistivity has the temperature dependence
of an insulator $(d\rho/dT<0)$ rather than a metal $(d\rho/dT>0)$.
The investigations presented in this paper are an extension of
these studies to the non-superconducting region of the $H-T$ phase
diagram. We used a 28~T resistive magnet at the Grenoble High
Magnetic Field Laboratory, in order to measure the in-plane and
out-of-plane magnetoresistance with various field orientations
relative to the $ab$-plane of the crystal.

\section{Experiment}

\begin{figure}
\includegraphics[width=0.6\linewidth,angle=0,clip]{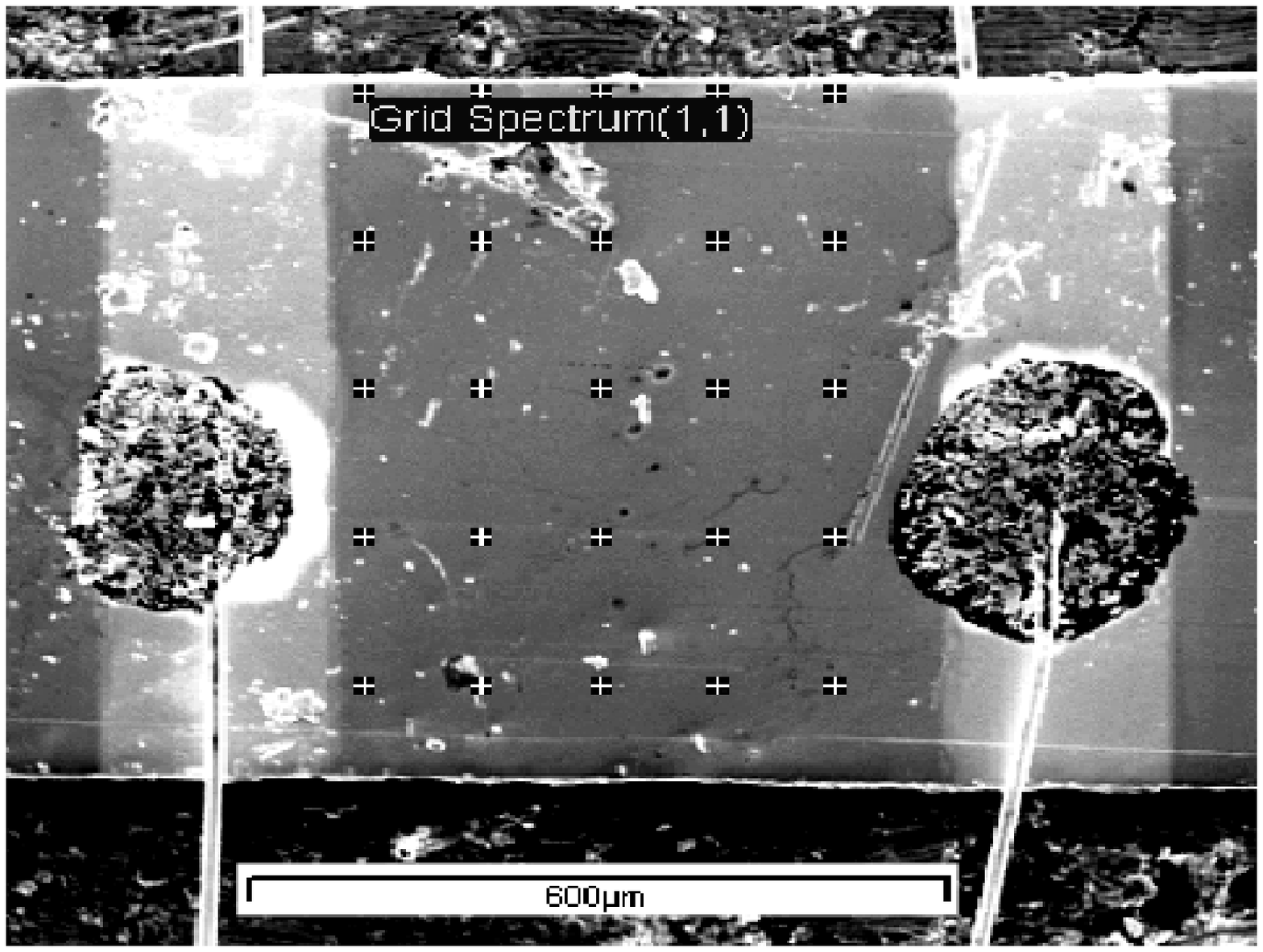}
\caption{\label{fig1} Scanning electron micrograph of the
investigated crystal between voltage contacts where the
composition measurement points are denoted by crosses.}
\end{figure}

The preparation and characterization of Bi2201 single crystals are
described in detail elsewhere.\cite{Vedeneev04} Here we have
characterized three \textit{as-grown} single crystals which have
been grown in the same crucible and found that samples are of high
quality with almost identical characteristics. To investigate the
low-temperature magneto-transport behavior, we selected the best
crystal. The crystal was cut to have the approximate dimensions of
$1.9$~mm $\times0.7$~mm$\times10~\mu$m. The actual cationic
composition of the selected sample was measured at 40 different
points on the crystal and the scatter in the data was less than
2\%. Figure~\ref{fig1} shows the scanning electron micrograph of
the crystal between the voltage contacts (gold film, a silver pad
and gold wires) where the composition measurement points are
denoted by crosses. We estimate the carrier concentration in the
sample to be $p=0.09$ by using the empirical relation between the
Bi excess, $x$, and $p$.\cite{Ando00,Vedeneev04} Optimum doping in
this system occurs around $p\simeq 0.17$ and the $T_c(p)$
dependence shows a faster drop in the underdoped side of the phase
diagram. The superconducting phase in B2201 extends only down to
$p=0.11$.\cite{Vedeneev04} The error associated with the carrier
concentration estimation is less than 4\% (see Fig.\ref{fig1} in
Ref. [\onlinecite{Bel}]). Thus, our sample with $p=0.09$ is
definitely located below the superconductor-insulator phase
transition, heavily underdoped and
non-superconducting.\cite{Vedeneev04} The half-width of the
sublattice reflections in the X-ray rocking curves for this
crystal did not exceed $0.15^{\circ}$. These data clearly
demonstrate the high structural quality and high homogeneity of
the sample on a microscopic scale.

A four-probe contact configuration, with symmetrical positions of
the low-resistance contacts ($<1\Omega $) on both $ab$-surfaces of
the sample was used for the measurements of $R_{ab}$ and $R_c$
resistances. The temperature and magnetic field dependence of the
resistances $R_{ab}(T,H)$ and $R_c(T,H)$ were measured using a
lock-in amplifier driven at $\approx $10.7 Hz. For the low
temperature magnetotransport measurements, the crystal was placed
directly inside the mixing chamber of a top-loading dilution
fridge. For the in-plane transport current $\mathbf{J}$, a
configuration with $\mathbf{H\perp J}$ was used in all cases. In
the angular magnetoresistance measurements for the out-of-plane
transport current, the magnetic field direction changed from the
longitudinal ($\mathbf{H\parallel c\parallel J}$) to transverse
($\mathbf{H\perp c\parallel J}$) configurations. The angular
resolution was better than $0.3^{\circ}$. The ac current applied
was $5 \mu$A for in-plane and $10 \mu$A for out-of-plane
resistance measurements. A RuO$_2$ thermometer was used to measure
the local temperature of the sample. The field sweep rate $dH/dt =
0.5$~T/min at temperatures $30-150$~mK and $1$~T/min at higher
temperatures was chosen in order to avoid eddy current heating.
The temperature was continuously recorded during each measurement
sweep.

\section{ In-plane [$\rho_{ab}(T)$] and out-of-plane [$\rho_{c}(T)$] resistivities}

\begin{figure}
\includegraphics[width=0.6\linewidth,angle=0,clip]{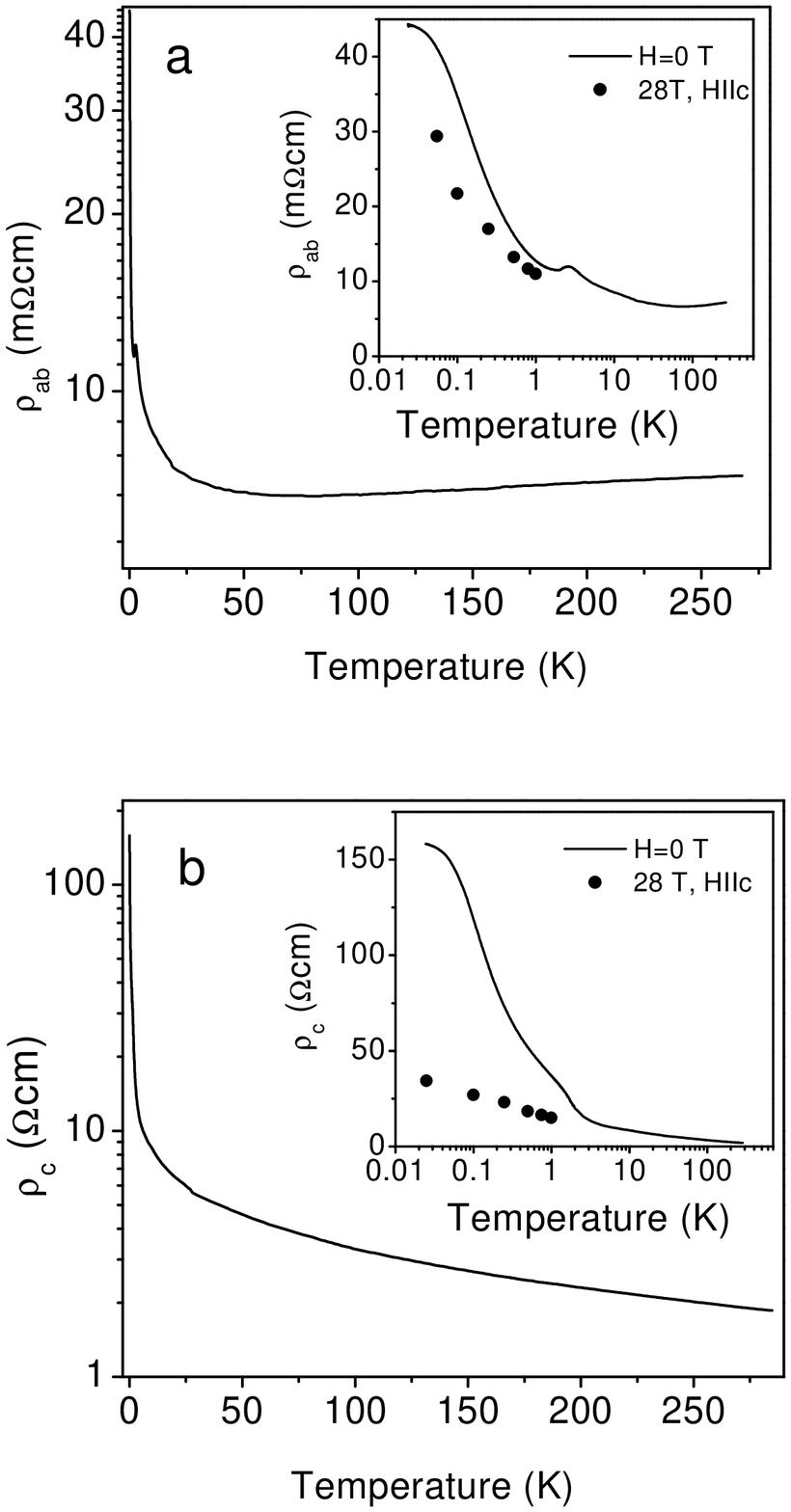}
\caption{\label{fig2} Temperature dependence of the in-plane
$\rho_{ab}$ (a) and out-of-plane $\rho_{c}$ (b) resistivities for
the investigated Bi2201 single crystal. Note the log scale for the
vertical axis. The insets plot $\rho_{ab}(T)$ and $\rho_c(T)$ with
the horizontal axis plotted using a logarithmic scale in order to
emphasize the low-temperature behavior.}
\end{figure}

In Fig.\ref{fig2} (main panels) we show the temperature dependence
of the in-plane $\rho_{ab}$ (a) and out-of-plane $\rho_{c}$ (b)
resistivities for the investigated Bi2201 single crystal, with the
vertical axis plotted using a logarithmic scale. The insets in
Fig.\ref{fig2} plot $\rho_{ab}(T)$ and $\rho_c(T)$ with a
logarithmic scale for the horizontal axis in order to emphasize
the low-temperature behavior. The data points (closed circles)
show the resistivity data at $H=28$~ T with the magnetic field
parallel to the $c$-axis. Figure~\ref{fig2} clearly demonstrates
that at zero magnetic field, the sample remains in its a normal
state down to 20 mK. $\rho_{ab}$ in the high-temperature range
shows a weak metallic behavior ($d\rho_{ab}/dT>0$), goes through a
minimum at temperature $T\approx 70$ K and then shows an
insulating behavior, consistent with the onset of localization.
$\rho_{c}(T)$ in Fig.\ref{fig2}(b) increases as $\log (1/T)$ as
the temperature decreases from the room temperature down to
$T\approx 5$~K and then transforms to an insulating behavior. At
ultra low temperatures, $T=0.02-0.1$ K, $\rho_{ab}$ and $\rho_{c}$
show a downward deviation (saturation) from the insulating
behavior. Such deviation from a $\log(1/T)$ dependence of the
in-plane and out-of-plane resistance of Bi2201 in normal state at
ultra low temperatures has been studied in detail in Ref.
[\onlinecite{Vedeneev04}]. As can be seen in Fig.\ref{fig2}(a), at
zero magnetic field, there is the weak upturn in the region $2-3$
K, which in reference [\onlinecite{Vedeneev04}], was attributed to
a competition between the onset of superconductivity and
localization. A weak feature in this temperature region is also
observed in $\rho_{c}(T)$ (Fig.\ref{fig2}(b)).

A strong 28~T magnetic field in the perpendicular geometry barely
suppresses the localization and therefore, the insulating behavior
of $\rho_{ab}$ persists [Fig.\ref{fig2}(a)] as in the case of
underdoped crystals in Ref. [\onlinecite{ Vedeneev04}]. Whereas
the effect of the high magnetic field on $\rho_{c}$ is very
noticeable. The insulating behavior of the $\rho _c(T)$ dependence
at low temperatures is significantly suppressed and $\rho _c(T)$
shows an almost identical $\log (1/T)$ dependence over the whole
temperature range [Fig.\ref{fig2}(b)] that can be interpreted as
the magnetic-field induced suppression of localization. The
behavior of the resistivities  described above is in agreement
with our previous results for Bi2201 single crystals in the
magnetic-field induced normal state \cite{Vedeneev04} and
completes the crossover picture from a metallic to an insulating
behavior of $\rho _{ab}(T)$ and $\rho _{c}(T)$ with decreasing
hole concentration.

\section{In-plane magnetoresistivity [$\rho_{ab}(H)$] }

\begin{figure}
\includegraphics[width=0.6\linewidth,angle=0,clip]{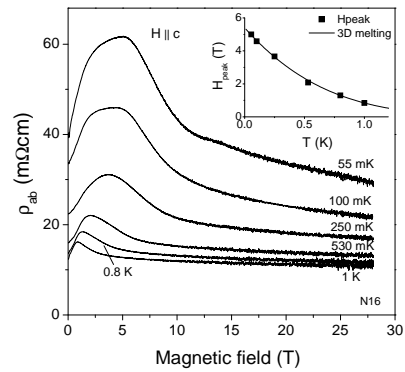}
\caption{\label{fig3} Transverse in-plane magnetoresistance for
Bi2201 sample at various temperatures from $55$ mK to $1$ K with
the magnetic field perpendicular to the $ab$-plane.
The inset is the magnetic field position of the maxima
$\rho_{ab}(H)$ (squares) versus the temperature.
The solid line
is a fit to the experimental points using the expression (1) in
Ref. [\onlinecite{Morello}] for an irreversibility line of our
low-$T_c$ Bi2201 single crystals that corresponds to the melting
of a three-dimensional vortex lattice (see text below). }
\end{figure}

Returning now to the magnetoresistance curves, we immediately
encounter an anomaly. Figure~\ref{fig3} (main panel) displays the
transverse in-plane magnetoresistance $\rho_{ab}(H)$ for the same
Bi2201 sample at various temperatures from $55$ mK to $1$ K with
the magnetic field perpendicular to the $ab$-plane. At each
temperature, the magnetic field dependence shows a crossover from
positive magnetoresistance at low magnetic field to negative
magnetoresistance at higher magnetic fields. As can be seen, the
resistivity starts to rise rapidly from the zero-field value,
reaches a maximum at the peak field, $H_{peak}$, and then
decreases sharply with increasing magnetic field. Thereafter, the
slope $d \rho_{ab}/dH$ changes and becomes small at high field.
This behavior points clearly to the presence of two different
mechanisms responsible for the negative magnetoresistance in
Fig.~\ref{fig3}. With increasing temperature, the maximum in
$\rho_{ab}(H)$ shifts toward lower magnetic field and the
amplitude of the maximum decreases monotonically. It is
significant that at low temperature (55~mK) the relative variation
$\Delta \rho_{ab} / \rho_{ab0} = [\rho_{ab}(H,T) - \rho_{ab}(0,T)
] / \rho_{ab}(0,T)$ is as much as +55\% at the peak field
(positive magnetoresistance) and -30\% at 28 T (negative
magnetoresistance) in the transverse configuration of the magnetic
field. It should be especially emphasized that the sharp change in
the slope $d \rho_{ab}/dH$ in the isotherms occurs at the
resistivities near the zero-field values.

For high-$T_c$ superconductors, a small positive transverse and
quadratic in-plane magnetoresistance has been observed in
Tl$_2$Ba$_2$CuO$_{6 +\delta}$ (3\% at 60~T and 130~K, $T_c \sim
80$~K) \cite{Tyler} and Bi2201 (12\% at 8~T and 5~K,
$T_c=3-4$~K).\cite{Vedeneev00a} However, $\rho _{ab}$ in
Fig.\ref{fig3} shows a much stronger positive magnetoresistance
compared to reported experimental results and $\Delta \rho_{ab} /
\rho_{ab0}$ does not show a quadratic dependence on magnetic
field. In quasi-classical models of conventional metals a
spin-dependent scattering leads to a very small ($\Delta \rho /
\rho \sim 10^{-3}$) positive magnetoresistance. It is known also
that the spin-dependent scattering leads to a positive
magnetoresistance that is independent of the applied field
orientation with respect to the current direction. Thus, \textit{a
priori}, the transverse magnetoresistance, in addition to the
orbital contribution, may also contain a Zeeman contribution
(actually, this is probably not the case, see below).

\begin{figure}
\includegraphics[width=0.6\linewidth,angle=0,clip]{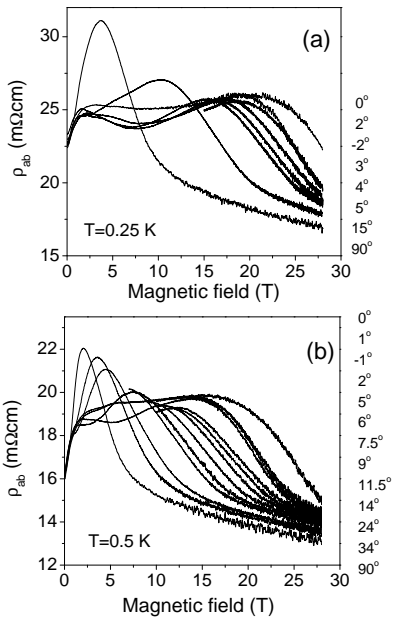}
\caption{\label{fig4} In-plane resistivity as a function of
applied field for various magnetic field orientations relative to
the $ab$-plane of the crystal, measured at 0.25 (a) and 0.5~K
(b).}
\end{figure}

To understand the origin of the orbital contribution to the
transverse in-plane magnetoresistance in our sample, we have
studied the angular dependence of the in-plane magnetoresistance.
Figure~\ref{fig4} displays the in-plane resistivity as a function
of applied field for various magnetic field orientations relative
to the $ab$-plane of the crystal, measured at 0.25 (a) and 0.5~K
(b). We can clearly see a considerable difference in the field
position of the maximum in $\rho _{ab}(H)$ between the
perpendicular ($\theta=90^{\circ}$) and parallel
($\theta=0^{\circ}$) configurations of the magnetic field. Since,
in both field directions, the magnetoresistance is transverse, the
highly anisotropic response points to the importance of orbital
effects for the magnetoresistance. This probably excludes any
explanation of the positive in-plane magnetoresistance in terms of
spin effects. However the large anisotropy is restricted to the
region of the peak (the bell-shaped part of the curves) in $\rho
_{ab}(H)$. As can be seen from Fig.\ref{fig4}, at higher magnetic
fields, where $\rho _{ab}(H)$ shows a small negative
magnetoresistance, the curves remain almost unchanged as the
sample is rotated. Most likely, the high magnetic field region for
which $d \rho_{ab}/dH$ is small and negative in Fig.\ref{fig3} and
Fig.\ref{fig4}, corresponds to the negative transverse in-plane
magnetoresistance reported in earlier investigations of
non-superconducting Bi2201 single crystals.\cite{Jing}

The previously reported negative in-plane magnetoresistance was
quite small (near 5~\% at 0.45~K and 8~T).\cite{Jing} The
magnetoresistance was negative in the temperature range 0.45-20 K
and became positive with increasing temperature above 20 K. To
account for these results, the authors invoked localization
theory,\cite{Lee} which describes the low-temperature negative
magnetoresistance in metals in a weak-localization regime. The
negative magnetoresistance resulted from the magnetic field
induced suppression of localization effects. In zero magnetic
field, samples showed an insulating behavior of $\rho_{ab}(T)$ for
$T<20$~K, where localization effects should be important,
especially at very low temperatures. Although the negative
in-plane magnetoresistance observed at high magnetic fields in
Fig.\ref{fig3} is considerably larger than that observed in Ref.
[\onlinecite{Jing}] (due to ultra low temperatures), our results
are in agreement with these experiments and it is reasonable to
assume that they have the same physical origin.

\begin{figure}
\includegraphics[width=0.6\linewidth,angle=0,clip]{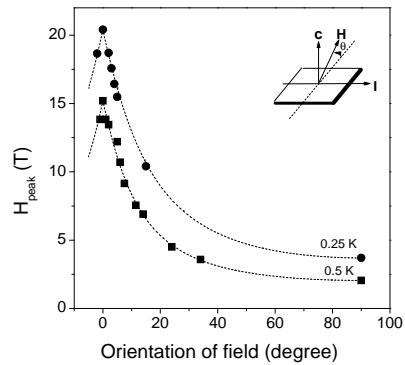}
\caption{\label{fig5} Angular dependence of the field position of
the maximum of $\rho _{ab}(H)$ ($H_{peak}$) at the temperatures
0.25 and 0.5~ K extracted from the magnetoresistance curves in
Fig.\ref{fig4}. Dashed lines are fits to experimental data using
Tinkham's relation with anisotropy parameter $\gamma =
H_{c2}^{\parallel ab} / H_{c2}^{\perp ab}$ equals to 5.5 at
$T=0.25$~K and 7.4 at $T=0.5$~K.}
\end{figure}

Turning now our attention to the origin of the bell-shaped part of
magnetoresistance, we would like to note that in layered
superconductors, the anisotropy of the magnetoresistance is a
direct consequence of the anisotropy of the upper critical field
$H_{c2}$. In such superconductors with a high degree of
anisotropy, a two-dimensional situation with decoupled layers
arises and for the angular dependence of $H_{c2}$ it is possible
to use Tinkham's relation for a thin-film superconductor in the
vicinity of $T_c$, \cite{Tinkham}
\begin{equation}
\mid H_{c2}(\theta )\sin\theta/H_{c2}^{\perp ab}\mid +
[H_{c2}(\theta )\cos{\theta}/H_{c2}^{\parallel ab}]^2=1.
\end{equation}

In Fig.\ref{fig5}, we show the angular dependence of magnetic
field position of the maximum of $\rho _{ab}$ at the temperatures
0.25 and 0.5~ K extracted from the magnetoresistance of the
crystal in Fig.\ref{fig4}. Dashed lines are fits to the data using
Tinkham's relation, with anisotropy parameter $\gamma =
H_{c2}^{\parallel ab} / H_{c2}^{\perp ab}$ equal to 5.5 at
$T=0.25$~K, and 7.4 at $T=0.5$~K. As can be verified from
Fig.\ref{fig5}, the magnitudes 5.5 and 7.4 are simply equal to the
ratio of the field position of the $\rho_{ab}(H)$ maxima at
$\theta=0^{\circ}$ and $\theta=90^{\circ}$. The experimental
points in Fig.\ref{fig5}, show a cusp-like behavior at
$\theta=0^{\circ}$ with $dH_{c2}/d\theta \neq 0$, in good
agreement with the prediction of the thin-film model (dashed
lines). This cusp-like behavior has previously been observed in
superconducting multilayers.\cite{Jin,Ghosh}

\begin{figure}
\includegraphics[width=0.6\linewidth,angle=0,clip]{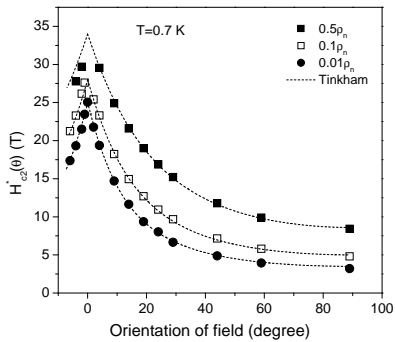}
\caption{\label{fig6} The resistive upper critical field
$H^*_{c2}$ as a function of the angle between the magnetic field
and $ab$-plane at $T=0.7$~K extracted from the magnetic
field-induced transitions of our superconducting slightly
underdoped Bi2201 single crystal with $T_c=6-7.5$~K. The different
symbols are the experimental $H_{c2}^{\ast }$ values obtained from
the fields at which the resistivity of the sample has reached 1\%,
10\% and 50\% of it normal-state value $\rho _{n}$.}
\end{figure}

Although Tinkham's model predicts a temperature-independent
critical-field anisotropy, a temperature dependence is
experimentally observed in layered superconducting single crystals
.\cite{Vedeneev02,Samuely, Ghosh} For comparison, we show in
Fig.\ref{fig6} the resistive upper critical field $H^*_{c2}$ as a
function of the angle between the magnetic field and $ab$-plane at
$T=0.7$~K extracted from the magnetic field-induced transitions of
one of our superconducting slightly underdoped Bi2201 single
crystals with $T_c=6- 7.5$~K. The different symbols are the
experimental $H_{c2}^{\ast }$ values obtained from the fields at
which the resistivity of the sample has reached 1\%, 10\% and 50\%
of it normal-state value $\rho _{n}$. Some data points are
missing, because of the very large values of the upper critical
fields in the ${\bf H\Vert {ab}}$ geometry ($\theta=0^{\circ}$) at
this temperature, we are unable to determine $H_{c2}^{\ast}$
values using the 50\% normal-state resistivity criterium. Again,
the experimental data, which shows a cusp-like behavior at
$\theta=0$, is in good agreement with the prediction of Tinkham's
formula (dashed lines).

In our opinion, Figs.\ref{fig3} - \ref{fig6} demonstrate clearly
that the observed positive in-plane magnetoresistance in $\rho
_{ab}$ (Fig.\ref{fig3}) is associated with superconductivity and
the maximum in $\rho _{ab}(H)$ at low temperatures is close to
some ``upper critical field''. However admittedly, the explanation
of the maximum itself in $\rho _{ab}(H)$, in particular the
bell-shaped form of the magnetoresistance curves in Fig.\ref{fig3}
is not fully understood.

In the inset of Fig.\ref{fig3} we show the magnetic field position
of the maxima $\rho_{ab}(H)$ (squares) versus the temperature. The
solid line is a fit to the experimental data using expression (1)
in Ref. [\onlinecite{Morello}] for the irreversibility line of our
low-$T_c$ superconducting Bi2201 single crystals which corresponds
to the 3D-2D melting of a vortex lattice. Here we have used the
fact that at very low temperatures the melting field and the upper
critical field coincide \cite{Shibauchi}. The fit shown in the
inset of Fig.\ref{fig3} is made fixing $T_c = 2.45$~K (the onset
of the weak upturn in Fig.\ref{fig2}(a)) and leaving $H^*_{c2}(0)$
and $c^2_L\sqrt{\beta_m/Gi}$ as free parameters. Here $c_L$,
$\beta_m\approx 5.6$ and $Gi$ are the Lindemann number, a
numerical factor and the Ginzburg number, respectively.
\cite{Morello} From the fit we obtain $H^*_{c2}(0)=5.4$~T and
$c^2_L\sqrt{\beta_m/Gi}=0.37$. The value of $H^*_{c2}(0)$ is close
to the experimental magnitude $H_{peak}$=5~T at $T=55$~mK and
$c^2_L\sqrt{\beta_m/Gi}$ closely matches the value found in Ref.
[\onlinecite{Morello}] taking into account the difference in
``$H^*_{c2}(0)$''. This is indicative of the presence of a vortex
state in the non-superconducting sample in magnetic field. In this
case, the maximum and negative magnetoresistance in the
bell-shaped part of $\rho _{ab}(H)$ in Fig.\ref{fig3} can be
explained by the fact that at high magnetic fields, the number of
regions able to support vortices decreases with increasing
magnetic field, and therefore, the dissipation decreases.

A linear extrapolation to high temperatures of the amplitude of
the maximum $\rho_{ab}$ plotted on a log scale as a function of
temperature (not shown) from Fig.\ref{fig3}, shows that the
vortex-like excitations have to vanish at a critical temperature
$T_{\phi}\approx 4-5$~K. Hence, using $T_{c}=2.45$~K for our
sample gives $1.5 T_{c} < T_{\phi} < 2 T_{c}$. This is consistent
with the existence of vortex-like excitations in the pseudogap
region, up to a temperature $T_{\phi}$, that manifest themselves
as flux-flow resistivity.\cite{Sandu}

Another possible explanation for the bell-shaped magnetoresistance
in Fig.\ref{fig3}, suggested by the quasi-two-dimensional nature
of Bi2201, is a magnetic-field-tuned superconductor-insulator
transition. Very recently Steiner $et~al.$ \cite{Steiner}
suggested that the behavior of some high-$T_c$  superconductors at
very high magnetic fields is similar to that of amorphous indium
oxide (InOx) films near the magnetic field tuned
superconductor-insulator transition. A similar magnetoresistance
peak at low temperatures was first observed by Paalanen $et~al.$
\cite{Paalanen} in amorphous superconducting InOx thin films. The
authors of Ref.[\onlinecite{Paalanen}] explained this peak by
invoking the scaling theory of the superconductor-insulator
transition in disorder two-dimensional
superconductors.\cite{Fisher}

In such systems, at sufficiently low magnetic fields and at zero
temperature, Cooper pairs are condensed, whereas field-induced
vortices can be localized due to disorder (a vortex glass phase).
As the magnetic field is increased, the system undergoes a
superconductor-insulator transition at some critical field. In the
insulating phase, the vortices are delocalized and undergo a
condensation, whereas Cooper pairs are localized. Near the
transition, there is a competition between condensation of Cooper
pairs and vortices. In the vortex-glass phase, long-range
crystalline correlations are destroyed by disorder vortex motion
at finite temperatures, giving rise to a nonzero resistance.
Further increasing the magnetic field causes the system to enter a
Fermi-insulator state containing localized single electrons.
Experimental evidence for this picture has been provided by the
temperature and magnetic-field dependence of the resistance in
amorphous InOx films.\cite{Paalanen} In order to explain the
observed peak in magnetoresistance near the
superconductor-insulator transition and the subsequent decrease of
the magnetoresistance, it has been suggested that an insulator
with localized Cooper pairs should have a higher resistance than
an insulator with localized single electrons.

Steiner $et~al.$ \cite{Steiner}  argued that a local pairing
amplitude persists well into the dissipative state of the
high-$T_c$ superconductors, the regime commonly denoted as the
``normal state'' in very high magnetic field experiments. They
concluded that the superconductor-insulator transition in
La$_{2-x}$Sr$_x$CuO$_4$ occurs at a critical field of the order of
the mean-field upper critical field $H_{c2}(0)$ where the
magnetoresistance maximum at low temperatures was
observed.\cite{Ando95} They attributed the magnetoresistance
maximum to a decrease in the local pair amplitude and a crossover
from a Bose-particle dominated to a Fermi-particle dominated
system. This means that the large resistance of the sample is
dominated by weakly localized pairs, while above the maximum for
$H > H_{c2}(0)$, pairs start to dissociate at a faster rate,
giving rise to a negative magnetoresistance as the system slowly
approaches a state that does not support pairing.\cite{Steiner}

The magnetoresistance curves presented here in Fig.\ref{fig3} are
not strictly identical to those reported for amorphous
superconducting InOx films\cite{Paalanen,Steiner} in which all
isotherms reach a maximum and then start to decrease at a common
magnetic field value. In our sample the magnetic field position of
the maximum magnetoresistance changes with the temperature. In
addition, the resistance of InOx films beyond the maximum
saturates at high magnetic field but remains larger by a factor of
$1.7$ than the zero-field normal-state resistance, as extrapolated
from the temperature dependent resistance above the transition.
Whereas in our case, the magnetoresistance in the bell-shaped part
of the curves reaches the zero-field normal-state value and
further decreases at high magnetic field (Fig.\ref{fig3}).
Nevertheless, the behavior of our strong disordering
non-superconducting Bi2201 sample in the low and moderate magnetic
fields with the bell-shaped magnetoresistance is similar to that
of thin films of amorphous InOx near the magnetic-field-tuned
superconductor-insulator transition.\cite{Paalanen,Steiner}. Our
interest is in the superconductor-insulator transition in a system
with a strong disorder which drives $T_c\rightarrow0$ so that bulk
superconductivity is suppressed. We note that the description of
the superconductor-insulator transition above does not take into
account the role of unpaired electrons, which are expected to
become important when the magnetic field is high enough to ensure
the destruction of the localized pairs. Here, as in Ref.
[\onlinecite{Steiner}], we think that the superconductor-insulator
transition in a strongly disordered Bi2201 system occurs close to
the mean-field upper critical field $H_{c2}(0)$ where the
magnetoresistance in Fig.\ref{fig3} has maximum.

\begin{figure}
\includegraphics[width=0.6\linewidth,angle=0,clip]{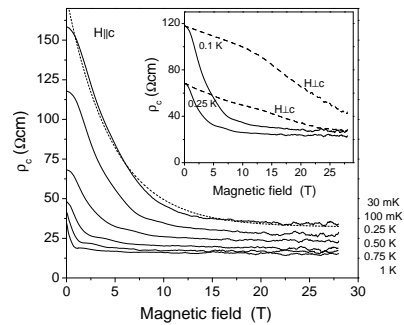}
\caption{\label{fig7} (Main panel) $\rho_c$ versus magnetic field for
Bi2201 single crystal at
various temperatures from $30$ mK to $1$ K for magnetic field
perpendicular to the $ab$-plane
($\mathbf{H\parallel c\parallel J}$). The inset is $\rho_c(H)$ curves at
temperatures 0.1 and 0.25~K for
the longitudinal $\mathbf{H\parallel c\parallel J}$ (solid lines) and
transverse $\mathbf{H\perp c\parallel J}$ (dashed lines) configurations. }
\end{figure}

\section{Out-of-plane magnetoresistivity [$\rho_{c}(H)$] }

In Fig.\ref{fig7} (main panel) we plot  $\rho _c$ versus magnetic
field for the same non-superconducting Bi2201 single crystal at
various temperatures from $30$ mK to $1$ K for magnetic field
perpendicular to the $ab$-plane ($\mathbf{H\parallel c\parallel
J}$). One can see that the longitudinal out-of-plane
magnetoresistance is negative over the entire magnetic field
region and decreases by almost 80\% by $H=28$~T for $T=30$~mK. The
negative magnetoresistance in Fig.\ref{fig7} rapidly weakens with
increasing magnetic field and clearly shows a saturation. These
results are surprising because in heavily underdoped
superconducting Bi2201 samples, with $p=0.12$ and 0.13
($T_c$(midpoint)=2.3 and 3~K, respectively), after the suppression
of superconductivity by magnetic field, $\rho _c$ decreases almost
linearly with increasing magnetic field by 10-15\% at 28~T and 40
mK.\cite{Vedeneev04}

It was found in Ref.[\onlinecite{Vedeneev04}] that such a behavior
of $\rho_c(H)$ is typical for slightly underdoped, optimally doped
and overdoped superconducting Bi2201 samples. Moreover, the
normal-state out-of-plane magnetoresistance of superconducting
samples was independent of the field orientation with respect to
the current direction.\cite{Vedeneev00b} Whereas, the out-of-plane
magnetoresistance in the non-superconducting sample is highly
anisotropic. To illustrate this, we show in the inset of
Fig.\ref{fig7} the $\rho_c(H)$ curves at temperatures 0.1 and
0.25~K for the longitudinal $\mathbf{H\parallel c\parallel J}$
(solid lines) and transverse $\mathbf{H\perp c\parallel J}$
(dashed lines) configurations. As in the case of the in-plane
magnetoresistance in Fig.\ref{fig4}, there is a large difference
between the out-of-plane magnetoresistance behavior for two
magnetic field orientations relative to the $ab$-plane of the
crystal. This is further evidence that the observed out-of-plane
magnetoresistance of the non-superconducting sample may be
associated with the superconductivity.

\begin{figure}
\includegraphics[width=0.6\linewidth,angle=0,clip]{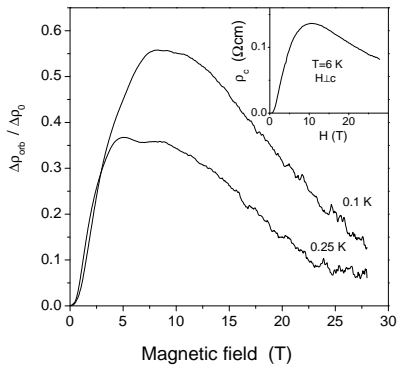}
\caption{\label{fig8} Orbital components of the transverse
out-of-plane magnetoresistance at temperatures 0.1 and 0.25~K
extracted from the data in the inset of Fig.\ref{fig7}. The inset
displays the field dependence of the out-of-plane resistivity
$\rho_c$ in the transverse configuration for our optimaly doped
Bi2201 single crystal with $T_c = 9.5$~K
(midpoint).\cite{Vedeneev00b}}
\end{figure}

As the response of a superconductor should be to the orbital
effect of a magnetic field, we assume following Jing \textit{et
al.}, \cite{Jing} that the longitudinal magnetoresistance involves
the spin degrees of freedom alone, and that these contributions
are isotropic. Then the orbital contribution to the transverse
magnetoresistance may be obtained by subtracting the longitudinal
magnetoresistance from the transverse, i.e. $\Delta\rho_{orb} =
\Delta\rho_T - \Delta\rho _L$. Here $\Delta\rho_{T,L} =
[\rho_{c}(H,T) - \rho_{c}(0,T) ] / \rho_{c}(0,T)$. In
Fig.\ref{fig8} we display the orbital components of the transverse
out-of-plane magnetoresistance at temperatures 0.1 and 0.25~K
extracted from the data in the inset of Fig.\ref{fig7}. The curves
in Fig.\ref{fig8} are remarkably similar to the broad interlayer
resistive transitions of the superconducting Bi2201 single
crystals in the transverse configuration of the magnetic field
($\mathbf{H\perp c\parallel J}$). For comparison, we show in the
inset of Fig.\ref{fig8} the field dependence of the out-of-plane
resistivity $\rho_c$ in the transverse configuration for our
optimaly doped Bi2201 single crystal with
$T_c$(midpoint)$=9.5$~K.\cite{Vedeneev00b}

Turning back to Fig.\ref{fig7} (main panel), we see that the
out-of-plane negative magnetoresistance at ultra low temperatures
rapidly saturates with a  characteristic exponential decrease with
magnetic field. The short dashed curve shows a numerical fit to
the magnetoresistance data at 30~mK calculated using the
functional form $\rho _c(H,T)=\rho_{c0}+a\exp (-H/bT)$, where $a$
and $b$ are constants. In previous measurements \cite{Vedeneev04}
we have pointed out that this Zeeman-like expression describes the
anisotropic negative out-of-plane magnetoresistance in the
superconducting Bi2201 samples, where, the major contribution to
the $\rho_c(H)$ curves is due to the gradual decrease of the
superconducting gap. If we suggest that a vestige of
superconductivity exists in the non-superconducting sample, then a
comparison of the temperature-dependent data in Fig.\ref{fig2}(b)
with the data in Fig.\ref{fig7}, allows us to conclude that the
observed negative magnetoresistance corresponds to a suppression
of the insulating behavior in $\rho _c(T)$, which can in turn be
interpreted as the magnetic-field induced suppression of the
superconducting gap. This is simply a consequence of the fact that
Cooper pairs exist at low temperatures in our non superconducting
sample, while bulk superconductivity is suppressed by strong
disorder.

The data in Fig.\ref{fig5} suggest that the ``depairing'' field
$H_{c2}^{\parallel ab}$ in the parallel ($\theta=0^{\circ}$)
configuration of the magnetic field at 0.25~K is 20.5~T. In
previous measurements \cite{Vedeneev99,Vedeneev04} we have shown
that the maximum in $\rho_c(H)$ for Bi2201 single crystals does
not coincide with $ H_{c2}$ and that is positioned near
$0.4\rho_{ab}^n$, where $\rho _{ab}^n $ is the normal-state
resistivity in $ab$-plane. Moreover, the $\rho_c(H)$ curves have a
pronounced break-point in the derivative well above the
$\rho_c(H)$ peak. The field position of these break-points in the
derivative coincide with the $H_{c2}$ values determined from the
$\rho _{ab}(H)$ curves.\cite{Vedeneev99} As can be seen in
Fig.\ref{fig8}, the orbital components of the transverse
out-of-plane magnetoresistance ($\theta=0^{\circ}$) at temperature
0.25~K also start to saturate at $H \simeq 22$~T. The fact that
``$H_{c2}^{\parallel ab}$ magnitudes'' found from the in-plane and
out-of- plane magnetoresistance are in close agreement further
supports the presence the vortex-like excitations in the heavily
underdoped non-superconducting Bi2201 sample.

Regarding the large negative longitudinal magnetoresistance we
note that an anomalously large negative longitudinal
magnetoresistance (almost 90~\% at 50~mK and 8~T) has been
observed previously in the transition metal dichalcogenides
\cite{Kobayashi} which also have a layered structure. These
compounds show a typical temperature dependence characteristic of
variable-range hopping between localized states. Fukuyama and
Yosida \cite{Fukuyama} have explained this phenomenon by
introducing Zeeman shifts for the Anderson localized states
leading to enhanced conductivity (exponential in $g \mu_B H/k_B
T$) with the energy levels of one spin component closer to the
mobility edge. Here the $g$ is the Land\'{e} $g$-factor and
$\mu_B$ is the Bohr magneton. Since our sample shows an insulating
behavior for $\rho_{ab}(T)$ [Fig.\ref{fig2}(a)], we cannot exclude
that the large negative longitudinal magnetoresistance in
Fig.\ref{fig7} has the same origin as in Ref.[\onlinecite{
Kobayashi}].

Thus, there are strong grounds to believe that the data above
indicates a link between superconductivity and the observed
magnetoresistance of the non-superconducting sample. We can
formally exclude macroscopic sample inhomogeneity, as the origin
of the observed phenomena because the crystal is of a very high
quality, judging from the magnetization measurements, composition
analysis and X-ray measurements. The composition of the crystal
was studied using a Philips CM-30 electron microscopy with a Link
analytical AN-95S energy dispersion X-ray spectrometer that
permitted to study sample across the whole thickness ($10 ~\mu$m).
Since the X-ray penetration depth in the X-ray diffraction
measurements was nearly 6.5~$\mu$m, the crystal was investigated
on both sides.  The rocking curve width, was also found to be
identical.

Taken together, the evidence suggest that at high magnetic field,
the sample, in an insulating state, is populated by vortices as
observed in superconducting cuprates at $T>T_c$ in a magnetic
field, for example, by the detection of a large Nernst
signal,\cite{Corson, Xu, Capan, Wang02, Wang03} angular
magnetoresistivity measurements,\cite{Sandu} and a torque
magnetometry.\cite{Wang05} In a strongly disordered superconductor
in zero magnetic field at very low temperatures, localization
effects are strong, long-range phase coherence is destroyed and
the superconductivity is suppressed. Nevertheless, a vestige of
superconductivity and delocalized vortex-like excitations exist in
a magnetic field.\cite{Xu} Vortex motion at finite temperatures
leads also to a nonzero positive magnetoresistance (Fig.\ref{fig3}
and Fig.\ref{fig4}). In the vicinity of the maximum in $\rho
_{ab}(H)$ a melting of the vortices occurs resulting in a
competition between the positive magnetoresistance and the
negative in-plane magnetoresistance that is due to delocalized
unpaired electrons. The negative magnetoresistance is also caused
by the decrease of the number of regions able to support vortices
with increasing the magnetic field. Since the pair amplitude of
localized Cooper pairs is very slowly suppressed out to high
magnetic fields, the system retains a vestige of superconductivity
at magnetic fields well above $H_{c2}$. \cite{Steiner} The system
slowly approaches a state that does not support pairing where the
bell-shaped part of $\rho _{ab}(H)$ curves in Fig.\ref{fig3} is
completed. Thereafter the negative in-plane magnetoresistance is
due to the magnetic field dependent localization effects only.

\section{Conclusion}

We have presented the temperature dependence for both the in-plane
$\rho _{ab}(T)$ and out-of-plane $\rho _c(T)$ resistivities and
magnetoresistivities $\rho _{ab}(H)$ and $\rho _c(H)$ in a high
quality non-superconducting (down to 20 mK) La-free
Bi$_{2+x}$Sr$_{2-x}$CuO$_{6+\delta }$ single crystal. The metallic
behavior of $\rho_{ab}(T)$ gradually changes to insulating
behavior with decreasing temperature consistent with the onset of
localization. $\rho_{c}(T)$ increases as $\log (1/T)$ as the
temperature decreases from room temperature down to $T\approx 5$~K
and then transforms to an insulating behavior down to $20$ mK also
due to localization. A strong 28~T-magnetic field in the
perpendicular geometry barely suppresses of the insulating
behavior of $\rho_{ab}$. Whereas magnetic field effectively
suppresses the insulating behavior in $\rho _c(T)$ at low
temperatures and $\rho _c(T)$ shows an almost identical $\log
(1/T)$ dependence over whole temperature range. Again, this can be
interpreted as the magnetic-field induced suppression of
localization. By measuring the angular dependencies of in-plane
and out-of-plane magnetoresistivities at temperatures from 1 K
down to $30$ mK, we have obtained evidence for the presence of
vortex-like excitations in a non-superconducting cuprate in the
insulating state. Similar vortex-like excitations have been
previously observed in superconducting cuprates at $T>T_c$ in
magnetic fields by the detection of a Nernst signal.

\begin{acknowledgments}
We thank V.P. Martovitskii and S.G. Chernook for the careful characterization of the
single crystals. This work has been partially supported by NATO
grant PST.CLG. 979896.
\end{acknowledgments}

\bibliography{NS2_Resub}
\end{document}